\documentclass{pinchcr}   

\title{Magnetic field-induced novel insulating phase in 2D superconductors}

\author{G. Sambandamurthy}
\affiliation{Department of Physics, University at Buffalo,\\
The State University of New York, Buffalo, NY 14260, USA.}
\author{N. Peter Armitage}
\affiliation{Department of Physics and Astronomy, \\
The Johns Hopkins University, Baltimore, MD 21218, USA.}

\authors{2}

\begin{document}

\maketitle

\dedication{Dedicated to.....}

\tableofcontents

\maintext

\part{Superconductor - Insulator transitions}

\chapter{Magnetic field induced novel insulating phase in 2D superconductors}
In a superconductor, a zero-resistance, quantum many-body state emerges at low temperatures ($T$'s) inhibiting all scattering of electrons that can produce resistance. On the other hand, in an insulator, the motion of electrons is greatly hindered rendering them unable to carry current in the $T$=0 limit. At low-$T$, and at zero magnetic field ($B$), superconducting thin-films have an immeasurably low resistivity ($\rho$). An application of $B$ perpendicular to the film's surface weakens superconductivity and it is believed that at a well-defined critical field ($B_c$), superconductivity disappears and an insulating behavior sets-in. This $B$-induced superconductor-insulator transition (SIT) in 2D, first observed by Hebard and Paalanen \cite{Hebard90}, has been the subject of a large number of both theoretical \cite{FisherPRL1,FisherPRL2,Fink94,Feigel1,Nandini,Meir07} and experimental \cite{Markovic98,White86,Yazdani95,Gant98,Butko01,Biel02,Murthy04,Steiner2,Crane06,Valles07,Baturina07,Sacepe08} studies. The SIT in 2D systems is theoretically considered within the framework of continuous quantum phase transitions \cite{Sondhi97,Sachdevbook}. Fisher and co-workers \cite{FisherPRL1,FisherPRL2} postulated a dual description of the SIT in which the superconducting and the insulating phases are caused by the condensation of Cooper-pairs and vortices, respectively, into a superfluid state. In such a picture, local superconductivity must persist beyond the transition. According to Fisher's theory \cite{FisherPRL1}, the transition is manifested by a change in the macroscopic vortex-state, and Cooper pairs will exist, albeit localized, in the insulating phase to support the formation of vortices.
While several experimental \cite{Hebard90,Markovic98,Yazdani95,Gant98,Murthy04,Chris02} and theoretical \cite{FisherPRL1,FisherPRL2} studies are supportive of this conjecture, others \cite{Valles92} correlate the SIT with the vanishing of the superconducting gap and amplitude of the order parameter, indicating that Cooper-pairs do not survive the transition to the insulator. In this chapter, we present data obtained from our DC and AC conductivity studies of disordered thin films of amorphous indium oxide (a:InO) that point to a possible relation between the conduction mechanisms in the superconducting and insulating phases in the presence of a magnetic field. 

\section{Magnetic field as the tuning parameter}
\begin{figure}
\begin{center}
\includegraphics[width=14cm]{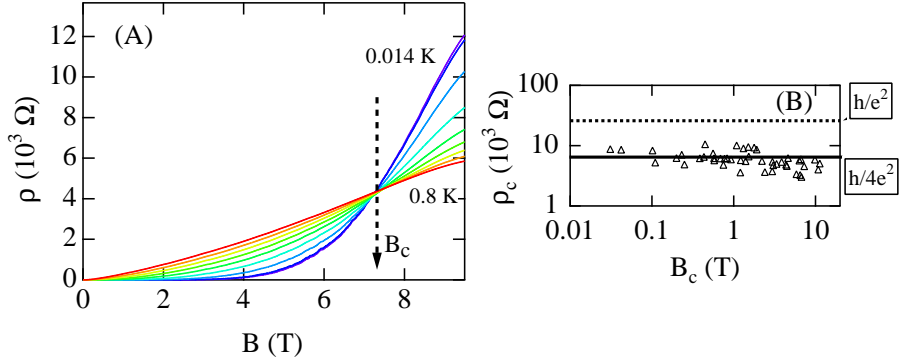}
\caption{(A) $\rho$ versus $B$ isotherms  at a low $B$ range for an a:InO film at $T$'s = 0.014 K, and 0.1 -- 0.8 K in 0.1 K interval. The critical point of the $B$-driven SIT, $B_c$, is indicated by the vertical arrow. (B) Critical resistance at the crossing point for our superconducting samples are plotted against the corresponding $B_c$ values. The solid horizontal line marks $h/4e^2$ and the dashed horizontal line marks $h/e^2$, respectively}.
\label{DC1}
\end{center}
\end{figure}

A typical magnetic field tuned superconductor-insulator transition in an a:InO film is demonstrated in Fig. \ref{DC1}. In Fig. \ref{DC1}(A)  $\rho$ {\it vs.}\ $B$ isotherms, taken at several $T$'s between 0.014--0.8 K, for a typical superconducting film are shown. Here, we focus on the lower $B$-range of the data, which includes the critical point of the $B$-driven SIT. The transition point is identified with the crossing point of the different isotherms at $B_c$ = 7.23 T. For $B<B_c$, the superconducting phase prevails with $\rho$ increasing with $T$, while for  $B>B_c$ an insulating behavior takes over and $\rho$ is a strongly decreasing function of $T$. For this sample, the value of $\rho$ at the transition is 4.3 k$\Omega$. The behavior in this low $B$-range is in accordance with previous observations of the $B$-driven SIT \cite{Hebard90,Markovic98,Yazdani95,Gant98,Biel02,Murthy04}. In the next section, the novel behavior in the insulating phase when the magnetic field is increased well beyond $B_c$ is discussed.

Hebard and Paalanen's results \cite{Hebard90} were in accord with Fisher's scaling theory argument \cite{FisherPRL1} which identified a universal sheet resistance separating a superconducting phase of localized vortices and Bose-condensed electron pairs from an insulating phase of Bose-condensed vortices and localized electron pairs . It is therefore interesting to note the resistance values at the critical point of the phase transition in our samples. 
In Figure \ref{DC1} (B) the $\rho$ value at $B_c$, $\rho_c$, from several superconducting a:InO samples that exhibited a well-defined crossing point are plotted. The data are scattered around $5.8$~k$\Omega$ and, with a standard deviation of $1.8$~k$\Omega$.
A scatter of about a factor of 3 in $\rho_{c}$ can not usually be taken as an indication of a universal number. However, if we consider the fact that the measured $\rho$'s themselves spans more than 10 orders of magnitude, a three-fold variation is not that large, indicating that the Cooper-pair quantum resistance $h/4e^{2}$ ($\approx$6.45 k$\Omega$) is of special significance for the transition. This value is in rough agreement with the theoretical prediction for this transition \cite{FisherPRL1}, and with other experimental studies in the literature \cite{Hebard90,Liu91}, with the notable exception of experiments done on MoGe films \cite{Yazdani95}. 

\section{Novel transport properties in the insulating phase}

\begin{figure}
\begin{center}
\includegraphics[width=6.7cm,angle=0]{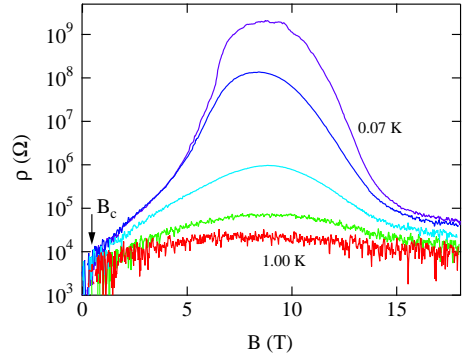}
\caption{$\rho$ versus $B$ isotherms at a large $B$ range for a sample at  $T$'s 0.07, 0.16 , 0.35, 0.62 and 1.00 K. The critical point of the $B$-driven SIT, $B_c$, is indicated by the vertical arrow.}
\label{DC2}
\end{center}
\end{figure}

The transport behavior of a superconducting film changes dramatically when we consider a broader range of $B$. In Fig. \ref{DC2}  isothermal curves from another sample, taken over a larger $B$-range, at several $T$'s between 0.07--1 K are presented.  As expected, when $B$ is increased beyond $B_c$ (0.45 T for this film), the insulating behavior initially becomes more pronounced. The surprising feature in our data is that, even though $\rho$ increases by more than five orders of magnitude from its value at $B_c$ (the ordinate in Fig. \ref{DC2} is plotted using a logarithmic scale) \cite{Butko01}, beyond approximately 9 T this trend reverses, the insulator becomes increasingly weaker and eventually $\rho$ approaches a value of 70 k$\Omega$ (at 0.07 K). This represents a drop of more than 4 orders of magnitude from the peak-value of 2 G$\Omega$. The demise of the insulating behavior seems to take place in two steps. An initial, rapid, drop beyond 10 T followed by a transition, at 14 T, to a slower decrease that continues all the way to, and beyond, our highest $B$ of 18 T. It is unclear at which $B$-value this trend will revert to the positive magnetoresistance expected at very high $B$.  First, we consider the details of the transport just above $B_c$ and closer to the resistance peak. We present three pieces of evidence that suggest a possible relation between the transport mechanism in the insulating and superconducting phases in a:InO films.

\subsection{Activation energy scales in the insulating phase}

\begin{figure}
\includegraphics[width=13cm,angle=0]{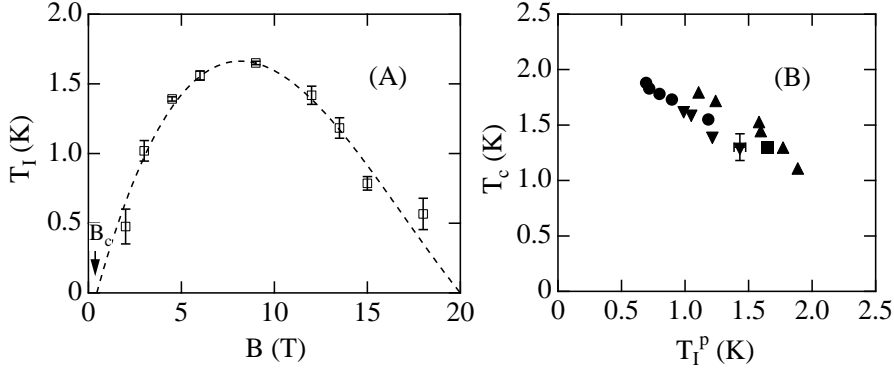}
\caption{(A) $T_I$, calculated from fits to Eq. 1, as a function of $B$ for sample in Fig. \ref{DC2}. $T_I$ has a peak at 9 T. The vertical arrow marks $B_c$ (= 0.45 T), where $T_I$ = 0. Dashed line is a guide to the eye.  (B) $T_c$'s at $B$ = 0 for several superconducting samples are plotted against $T_I$'s at the high-$B$ insulating peaks. Each physical sample is marked with a different symbol and samples have been annealed to vary $T_c$. Error bars indicated are typical of most data points.}
\label{DC3}
\end{figure}

Inspecting the details of the $T$ dependence of $\rho$ in the $B$-induced insulating phase reveals that the insulator follows an Arrhenius $T$-dependence,
\begin{equation}
\rho = \rho_0\ exp(T_I/T)
\end{equation}
indicating the existence of a mobility gap at the Fermi energy, with $k_BT_I$ the magnitude of the gap ($k_B$ is the Boltzmann's constant). Fitting our data to an Arrhenius form as in equation (1) yields $T_I(B)$, the activation temperature, that depends on $B$. In Fig. \ref{DC3} (A) we plot $T_I$ versus $B$ for the sample of Fig. \ref{DC2}. At $B_c$, the onset of the insulating behavior, $T_I$ is zero. It then increases with $B$ reaching a maximum at 9 T and, similar to $\rho$ itself, decreases at higher $B$. A rough extrapolation of our data indicates that $T_I$ will vanish at 20 T, although at $B$  values larger than $14$ T, the estimates of $T_I$ suffer from increasingly large errors, thus making this extrapolation inaccurate.

An interesting observation can be made when considering the values of $T_I$. At the insulating peak, $T_I$ = 1.65 K, conspicuously close to the $B$ = 0 value of $T_c$ for this film, 1.27 K. To test this observation we plot, in Fig. \ref{DC3} (B), $T_c$'s at $B$ = 0 for several of our superconducting samples against $T_I$'s at the insulating peak. Overall, $T_I$ is close to $T_c$, supporting the observation made above. However, in the cases where we are able to follow one physical sample through several anneal stages, we see that as $T_c$ increases upon lowering of the disorder, $T_I$ decreases, indicating that a more complete theory is required to account for the details of the $T_c$--$T_I$ dependence. The closeness of $T_I$ to $T_c$ suggests the existence of a relation, between the conduction mechanism of the superconductor and that of the insulator, on which more details are given below. Although the non-monotonic behavior of the $B$-induced insulator has been observed before in a:InO films \cite{Gant98,Paalanen92} and in other materials \cite{Butko01,Baturina07}, two new observations on this insulator require elucidation. They are: (a) the insulating peak in a:InO samples has an unusually large $\rho$ value and (b) the activation temperature of the insulator at the peak is close to the superconducting transition temperature of the film at $B$ = 0. In the reminder of this chapter, we present more evidence that suggests that it may be possible to account for both observations by adopting the point of view in which the properties of a:InO films are not uniform over the entire sample \cite{Nandini,Kowal94,Meyer01}.

\subsection{Current-voltage characteristics}

\begin{figure}
\includegraphics[width=13cm,angle=0]{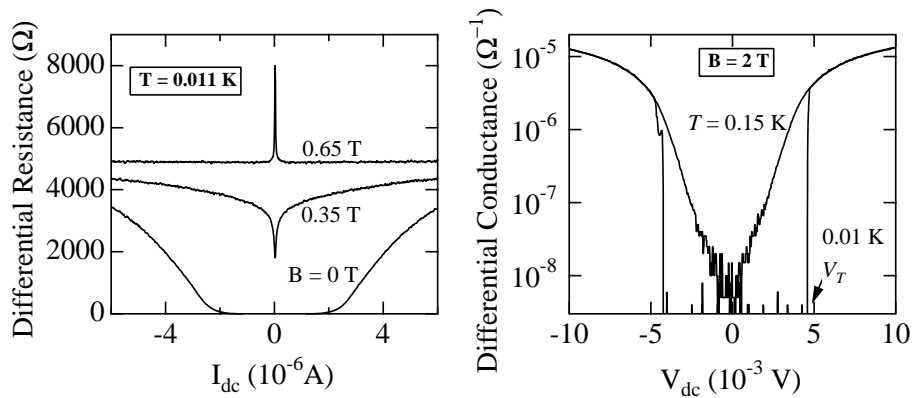}
\caption{Comparison of the current-voltage characteristics of the $B$-driven insulating phase at two $T$Õs (0.15 K and 0.01 K). The traces show the two-terminal differential conductance measured at $B$ = 2 T as a function of DC Voltage. The AC excitation voltage applied is 10 $\mu$V. The sample used is Ja5 with $B_c$ = 0.4 T. $V_T$ marks the threshold voltage for conduction at $T$= 0.01 K.}
\label{IV1}
\end{figure}

Several recent experiments, such as the ones presented above, provided strong evidence that at least some of the superconducting correlations remain in the insulating state of a:InO and other 2D superconducting films \cite{Murthy04,Valles07,Baturina07,Paalanen92,Gant00,Gant95}. We now turn to the current-voltage characteristics of the insulator, looking for more clues in support of the above observations. However, first it is instructive to examine the evolution of the non-linear current-voltage characteristics from the superconducting state through the transition and into the insulator. 

In Fig. \ref{IV1} (A) we plot the four-terminal differential resistance ($dV/dI$), as a function of the DC current ($I_{dc}$), taken from an a:InO film, measured at $T$ = 0.01 K. We show three representative traces that are measured in the $B$ = 0 -- 0.65 T range.  The bottom trace in Fig. \ref{IV1} (A), taken at $B$ = 0 T, is typical of a superconductor: $dV/dI$ is immeasurably low as long as $I_{dc}$ is below a well-defined value, $I_{dc}^c$, which is the critical current of the superconductor at that particular $B$-value. When $I_{dc} > I_{dc}^c$, superconductivity is destroyed and a dissipative state emerges. The situation is different for the middle trace of Fig. \ref{IV1} (A) taken at $B$ = 0.35 T. Even at $I_{dc}$ = 0 a zero-resistance state is not observed down to our lowest $T$. However, the current-voltage characteristic still maintains a superconducting flavor: it is non-Ohmic and $dV/dI$ increases with increasing $I_{dc}$. We therefore consider the $B$ = 0.35 T trace to represent a transitional state in the path the system takes from being a superconductor to an insulator. It is a matter of some debate whether this transitional state will develop into a full superconductor in the $T$ = 0 limit \cite{Yazdani95,Phillips03,Ephron96,Chervenak00}.  While the bottom two traces in Fig. \ref{IV1} (A) exhibit superconducting traits, the top trace taken at $B$ = 0.65 T clearly does not. Instead, it has the opposite low-$I_{dc}$ dependence indicative of an insulating state: it is again non-Ohmic, but this time an increase in $I_{dc}$ results in a decrease of $dV/dI$. This is consistent with the $\rho$-B data of this samples where the transition to the insulating phase occurs at $B_c$ = 0.4 T. While the non-linear current-voltage characteristic deep in the superconducting phase is usually attributed to current-induced vortex depinning \cite{Rzchowski90}, the origin of non-linearity in the insulating phase is less clear.

Let us now focus on the insulating phase well above $B_c$. Since we are now dealing with an insulator, it is natural to consider the differential conductance ($dI/dV$) rather than $dV/dI$. In Fig. \ref{IV1} (B) we plot two-terminal $dI/dV$ traces against the applied DC voltage ($V_{dc}$) measured at $B$ = 2 T, well above the $B_c$ (= 0.4 T) of this sample. In this Figure we contrast two traces that are measured at $T$ = 0.15 K and 0.01 K. The data taken at $T$ = 0.15 K are typical of an insulator, i.e., they are strongly non-Ohmic, having a low, but measurable, value at $V_{dc}$ = 0 that increases smoothly with increasing $V_{dc}$. No clear conduction threshold can be identified at this $T$. The response drastically changes when the film is cooled to 0.01 K: as long as $V_{dc}$ is below a well-defined threshold value ($V_T$), the value of $dI/dV$ is immeasurably low. At that $V_{dc}$ (= 4.65 mV for this sample) $dI/dV$ increases abruptly, by several orders of magnitude, and remains finite for higher values of $V_{dc}$. This clearly shows the emergence, at low $T$, of finite $V_T$ for conduction in the insulating phase.

\begin{figure}
\begin{center}
\includegraphics[width=8.7cm,angle=0]{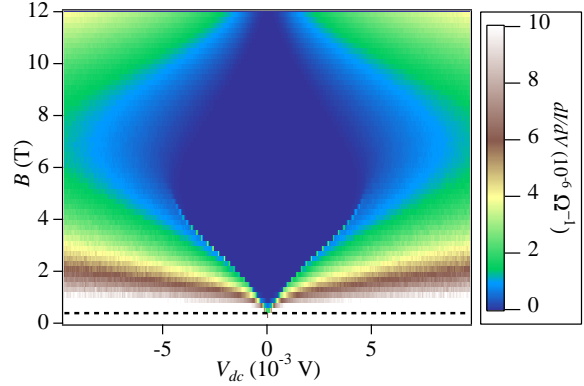}
\caption{Two-dimensional map of the $dI/dV$ values in the $B-V_{dc}$ plane. For the sample of Fig. \ref{IV1}, $dI/dV$ traces as a function of $V_{dc}$ at $B$ intervals of 0.2 T and at $T$ = 0.01 K are measured. The color scale legend on the right hand side shows the various colors used to represent the values of $dI/dV$. The horizontal dashed line denotes $B_c$ (= 0.4 T) of this sample.}
\label{IV2}
\end{center}
\end{figure}

We have so far described the observation of the emergence, at a well-defined temperature, of sharp thresholds for conduction on the insulating side of the SIT in a:InO films. Before the implications of this observation are discussed, it is illuminating to construct a map of the $B$-driven insulating phase in the $B$--$V_{dc}$ parameter space available in our experiments. In Fig. \ref{IV2} the behavior of a:InO films is summarized in the form of a 2D color map of the $dI/dV$ values in the $B$--$V_{dc}$ plane for the sample of Fig. \ref{IV1}. This map was constructed by measuring $dI/dV$ as a function of $V_{dc}$ at $B$ intervals of 0.2 T at 0.01 K. The colors in the map represent the values of $dI/dV$, changing from $dI/dV$ = 0 (dark blue) to $dI/dV$ = $10^{-5}$ $\Omega^{-1}$ (white). The horizontal dashed line marks $B_c$ (= 0.4 T) of this sample. Sharp conduction thresholds in the insulating phase can be seen in the map as a sudden change in color in the insulating regime between $B_c$ = 0.4 T and $\sim$ 5 T. As $B$ is increased the threshold behaviour appears at higher values of $V_{dc}$. This trend continues until $B$ is close to the $\rho$-peak, which is at $B$ = 6 T for this sample. Near the $\rho$-peak and beyond it, the $dI/dV$ traces no longer exhibit the sharp thresholds for conduction and $dI/dV$ increases smoothly with $V_{dc}$. This manifests as a gradual change in colours as $V_{dc}$ is changed for $B \geq$ 6 T. 

The sharp drop in $dI/dV$ values, at low $T$, may be evidence for the condensation of individual charge carriers into a collective state \cite{FisherPRL1,Chris02}. This brings about interesting analogies with a diverse class of physical systems showing similar threshold for conduction that are considered as signatures of collective phenomena. We recall two such examples here: the first is in one-dimensional organic and inorganic solids where thresholds to conduction have been treated as the signature of the depinning of charge density waves \cite{Gorkov89}. Similarly, in 2D electron systems confined in semiconductor heterostructures, thresholds for conduction have been attributed to the depinning of a magnetically induced electron solid akin to the Wigner crystal \cite{Goldman90,Jiang91,Williams91}. While the appearance, in disordered superconducting samples of a:InO, of a clear threshold voltage for conduction in the insulating phase can be considered as evidence for vortices condensing into a collective state at a well-defined temperature \cite{FisherPRL1}, more recent studies in a:InO and TiN films \cite{Ovadia09,Altshuler09,Vinokur10} have shown that the non-linear jumps in current-voltage characteristics result from a thermal decoupling of the electrons from the phonon bath in the $B$-induced insulating phase. This leads to interesting open questions about how such a phonon decoupling mechanism may arise.  Clearly more research is needed to understand this novel transport behavior in 2D superconductors. 

\subsection{Power law behavior in resistance}
The purpose of this section is to show that the resistivity of the superconducting a:InO films can be described by a single function covering a wide range of our measurements, which includes the $B$-driven SIT. This function can be written as follows:
\begin{equation}
\rho(B,T)=\rho_{c}(\frac{B}{B_c})^{T_{0}/2T}
\label{rlaw}
\end{equation}
where $\rho_{c}$, $B_c$ and $T_{0}$ are sample-specific parameters. The phenomenological form introduced above is consistent with the collective-pinning model of transport in thin superconducting films, which predicts a vortex-pinning energy proportional to $\ln(B)$ \cite{Blatter94a}. This form has been observed before in high-$T_c$ layered systems \cite{Palstra88,Inui89} as well as in amorphous superconductors \cite{Ephron96,White93,Chervenak96}. Here we show that this behavior is not restricted to the superconducting phase but continues, uninterrupted, well-into the insulating state\cite{Shahar98}.

\begin{figure}
\includegraphics[width=3.2in]{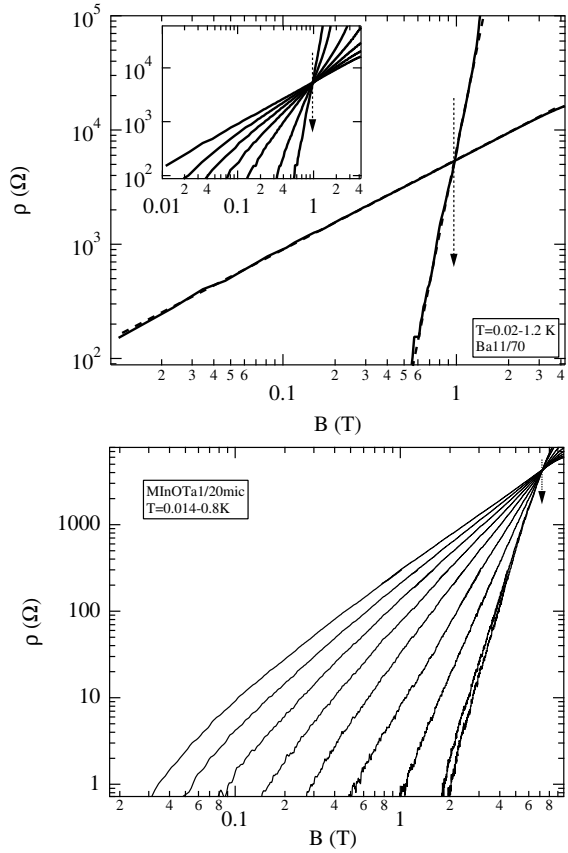}
\caption{ (A) $\rho$ versus $B$ isotherms for two of our samples presented in log-log graphs. Upper panel : Isotherms at $T$= 0.02 and 1.2~K are only shown for clarity. The inset shows the full set taken at $T$= 0.02~K, 0.2~K, 0.4~K, 0.6~K, 0.8~K,1.0~K and 1.2~K. Lower panel : Data from another sample taken at same $T$ values as in Fig.~1. Vertical arrows mark the crossing point of different $\rho$ isotherms. The $\rho$ isotherms display a power-law dependence on $B$, shown by the dotted lines in the main figure on the upper panel, that holds on either side of the crossing point.}
\label{powerLaw}
\end{figure}

We perform a quantitative analysis of the $B$ and $T$-dependence of the resistivity data, in the low $B$ superconducting regime and closer to $B_c$. In Figure \ref{powerLaw}  $\rho$ vs. $B$ traces taken at various $T$'s for two a:InO samples are plotted using log-log graphs. For the sample in the upper panel of Figure \ref{powerLaw}  special care is taken to extend the measurements over a large range of $\rho$. Each curve is well-described by a power-law dependence that holds over more than 2 orders of magnitude in $B$ and more than 3 in $\rho$, with non-random deviations that are only seen at high $T$s and as the $B$ values approach the insulating peak. The different curves are distinguished by their power, which is a function of $T$. A power-law $B$-dependence of $\rho$ in two-dimensional superconductors is associated with the collective-pinning flux-creep transport model, predicting an activation (pinning) energy, $U_0$, that depends logarithmically on $B$ \cite{Blatter94a}: 
\begin{equation}
U_0 \propto ln(B_c/B).
\label{pinE}
\end{equation}
This, in association with activated transport that can be described by 
\begin{equation}
\rho(B,T)=\rho_{c}e^{{-T_0(B)}/{T}}
\end{equation}
leads to the power-law dependence of Eq. \ref{rlaw}. Here we use $T_0$ as the sample-specific activation energy extracted from our data and $U_0$ as the activation energy as described in the collective pinning theory, though they imply the same physical meaning. Similar power-law behavior in $\rho$ was observed in disordered thin-films \cite{White93,Chervenak96,Ephron96} and in layered high-$T_c$ compounds \cite{Palstra88,Inui89}, and may be indicative of the central role played by vortices in our system. Reiterating the intriguing feature in our results we note that the power-law behavior described by Eq. \ref{rlaw} continues, uninterrupted, through $B_c$ and into the insulating state (see inset of Fig. \ref{powerLaw} (A)). Since, in the insulating phase, $B>B_c$ the pinning energy $U_0$ in Eq. \ref{pinE} becomes negative. This is a natural outcome of the physical picture described in \cite{FisherPRL1,FisherPRL2}, according to which the vortices, collectively pinned at $B<B_c$, condense at $B>B_c$ rendering the response of the system insulating. Consequently, each vortex that is thermally activated away from the condensate will cause a reduction in the resistance leading to the apparent negative $U_0$.

\section{Finite frequency behavior - superconducting correlations on the insulating side}
Complementary to the DC studies, finite frequency microwave measurements also reveal various aspects of the anomalous insulating state.  Using a custom low $T$ custom multimode cavity resonator, we studied the complex AC conductivity of thin disordered InO$_x$ films as a function of $B$ through the nominal SIT  \cite{Crane06,Crane06b}.   We resolved a significant finite
frequency superfluid stiffness well into the insulating regime.   These results can also be interpreted as evidence for collective effects and localized Cooper pairs on the superconducting side of the transition.

In general, AC measurements may provide a more complete characterization of low
energy properties of systems than DC measurements alone. The use
of finite frequencies allows various time and frequency scales to
be extracted.  Importantly in an AC measurement both the real and
imaginary parts of various response functions can be quantified also.
These can be, for example, the complex conductivity $\sigma_1 + i
\sigma_2$, the complex dielectric constant $\varepsilon_1 + i
\varepsilon_2$, or the complex sheet impedance of a thin film
$1/\sigma d = R + i \omega L$.  The ability to measure both the
real and imaginary components brings complementary information;
the real part of the conductivity (in-phase response) holds information about the
dissipative response of the system while the imaginary part (out-of-phase response) tells
something about the polarizability of charge carriers and/or their ability to move dissipationlessly. 

AC measurements as such are particularly useful in the characterization of superconductors.   For frequencies much less than the superconducting gap, the imaginary conductivity is
proportional to the superfluid density, which is related to the
`phase stiffness' as $\sigma_2 = \sigma_Q \frac{k_B
T_{\theta}}{\hbar \omega}$ where $T_{\theta}$ is the superfluid
stiffness in degrees Kelvin and $\sigma_Q $ is the quantum of
conductance for Cooper pairs divided by the sample thickness
$\frac{4e^2}{hd}$.  The superfluid stiffness is the energy scale for introducing phase slips in the superconducting order parameter $\Psi = \Delta e ^{i \phi}$ and is a fundamental quantity for understanding fluctuation
phenomena in superfluids and superconductors
\cite{Berez,KToriginal,KTBhelium,Hebard1985,Halperin,KTdynamics}.  Note that a number of different representations have been used in the literature to characterize the size of the out of phase AC response of superconductors (penetration depth, penetration depth squared, superfluid density, etc.)   In principle all of these are equivalent.   We prefer to use the phase stiffness as it is the most intrinsic quantity. 

$\sigma_2$ in a true long range ordered superconductor has a $1/\omega$ dependence that results from the Kramers-Kronig compatibility with the delta function at
$\omega=0$ in $\sigma_1$.  The Kramers-Kronig relations are integral relations between real and imaginary parts of response functions guaranteed by causality.  The coefficient of the $1/\omega$ is proportional to the superfluid stiffness, and because the quantity $\omega  \sigma_2$ is frequency independent in a long range ordered superconductor so is the superfluid stiffness.  When one has fluctuating superconductivity the delta function at $\omega = 0$ broadens and due to spectral weight conservation its moment must move to slightly higher frequencies.  

The fact that the Kramers-Kronig relation is an integral relation between $\sigma_1$ and $\sigma_2$ weighted by $1/\omega$ means that if this spectral weight in $\sigma_1$ is still concentrated narrowly enough at reasonably low frequencies, then one can still have a large $\sigma_2$. In general if $sigma_1$ is low and distributed over a broad spectral range as in a normal metal then $\sigma_2$ will be small.  Indeed with their low normal state conductivity and strong scattering, we cannot detect any significant out of phase component in our highly disordered thin films well above T$_c$. Thereby we can assign all of the enhancement of $\sigma_2$ that we detect to superconductivity.

In the case of a fluctuating superconductor where the delta function in $\sigma_1$ is removed, the $\sigma_2$ no longer has a strict $1/\omega$ dependence. However, if the spectral weight in $\sigma_1$ is concentrated at low frequency then $\sigma_2$ can still be appreciable.  If this enhanced $\sigma_1$ is due to fluctuating superconductivity, one can still define a generalized superfluid
stiffness, but now it has an explicit frequency dependence.  Our use of a frequency dependent stiffness is very similar to the treatment of the superfluid stiffness
in the context of the finite frequency measurements of the Kosterlitz Thouless
transition in liquid He4 films where one sees that the region of universal
discontinuous drop of the superfluid density broadens and shifts
to higher temperatures as the probing frequency is increased.  In
the context of Kosterlitz Thouless one considers that as the
frequency increases, one probes on shorter and shorter length
scales and thereby is less and less sensitive to the renormalizing
effects of intervening unbound vortex anti-vortex pairs. The expectation is that in the
high frequency limit the superfluid stiffness will behave
approximately mean-field-like and reflect the underlying
background superconductivity.  Aspects of this behavior has been observed in
thin granular Al layers \cite{Hebard80a}, thin indium/indium-oxide composites \cite{Fiory83a},
He4 thin films \cite{Adams87a}, as well as other related systems \cite{Corson}. 

Despite their potential impact there have been very few experiments using finite frequencies through the SIT.   This is largely due to the general difficulties and experimental constraints (the need for high, but not too high frequencies, low temperature, high magnetic fields) in performing such measurements.   In order to reveal the ground state fluctuation behavior, one generally needs $\hbar \omega \gg K_B T$, with $\hbar \omega < 2 \Delta$, where $\Delta$ is the superconducting gap.  This generally means that measurements must be performed in the microwave range (20 GHz corresponds approximately to 1 K) in devices like microwave cavities.  We performed such measurements in a novel cryomagnetic resonant microwave cavity system using the cavity perturbation technique.  In our case, the cavity diameter was optimized for performance in the 22 GHz ($\hbar \omega / k_B = 1.06$ K) TE011 mode.  However a number of other discrete frequencies from 9 to 106 GHz were accessible by insertion of an additional sapphire puck (for the low frequencies) or use of a very short "pan"-shaped cavity (for greater than 100 GHz). 

Our highest operating frequency of 106 GHz corresponds to an energy which is below the threshold for above gap excitation (158 GHz) in the sample studies here.  Relations between the cavity's resonance frequency shift $\Delta \omega$ and change in quality factor $\Delta (1/Q)$ upon sample introduction to the complex conductivity are obtained by a cavity perturbation technique
\cite{KotzlerBrandt1,KotzlerBrandt3,Redbook,Peligrad} (see
\cite{Waldron} for a very thorough treatment). This standard
experimental technique is based on the adiabatic modification of
the electromagnetic fields in a cavity that arises from the
introduction of a small sample to the interior of the cavity. Due to their thin film geometry, our
samples fall in the `depolarization regime in which the field penetrate the entire volume of the sample.  It has been shown that for extremely thin films, only in-plane AC
electric fields or out-of-plane AC magnetic fields at the sample
position can affect an appreciable change in a cavity's resonance
characteristics \cite{Peligrad2}. Samples were placed along the
cavity's central axis, where due to symmetry consideration and
depending on the particular TE mode being used, if the electric
field is in-plane then there is a zero out-of-plane magnetic field
and vice versa.  Modes with both these field configurations were exploited in our
setup and their analysis differed.  Please see Ref. \cite{Crane06b} for details.

\begin{figure}[htbp]
\begin{center}
\includegraphics[width=6.7cm,angle=0]{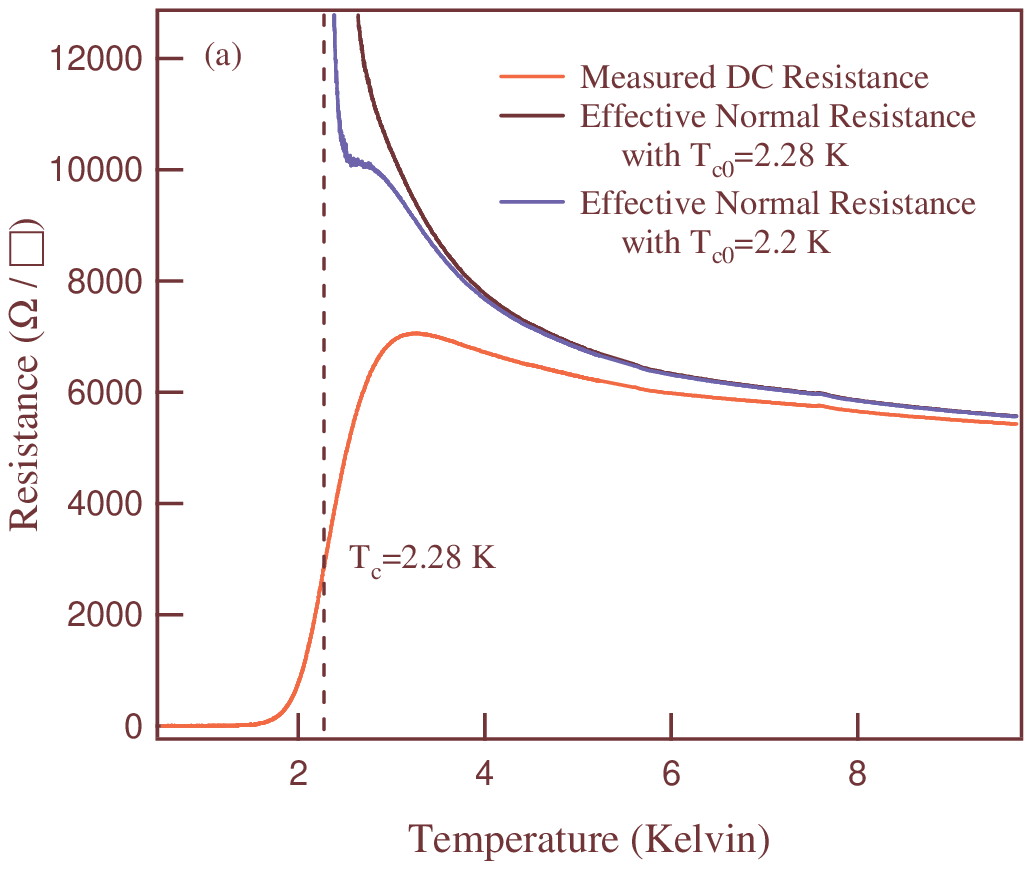}
\includegraphics[width=7cm,angle=0]{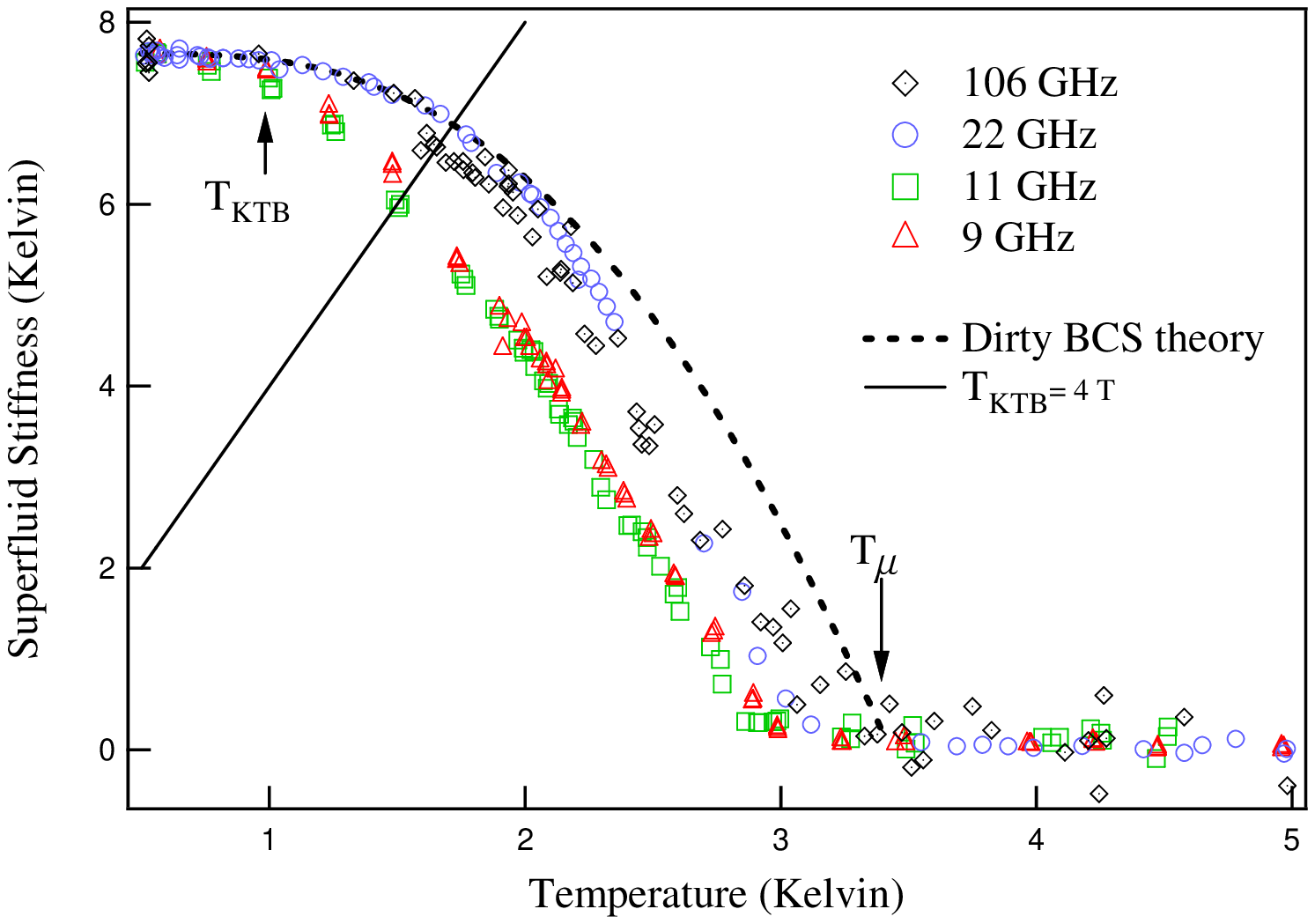}
\caption{ (top) DC Sheet Resistance. The normal state resistance curves are generated using the procedure described in the text.  T$_{co}$ is found to be 2.28 K.  (bottom) Temperature dependence of the superfluid stiffness $T_{\theta}$.  The prediction of the dirty BCS model is shown as a dashed line which goes to zero at T$_{\mu}$.  The lower frequency data show a significant deviation from the mean field behavior.  The temperature where $T_{\theta}$ acquires a frequency dependence can be identified with T$_{KTB}$. }
\label{AC1}
\end{center}
\end{figure}

We begin by first discussing the zero field behavior along the thermal axis of the field-temperature phase diagram.  Here we are interested in understanding the fluctuation behavior of the
complex order parameter $\Psi = \Delta e ^{i \phi}$ as the temperature is lowered towards the superconducting state.  The expectation is that there will be a temperature scale below which the amplitude of the order parameter will become defined, but the phase will still fluctuate.   At some lower temperature, the phase will order and a true superconducting state will occur.  In the DC data, as shown in Fig. \ref{AC1} (top), we observe a broad region over which the superconducting transition occurs.  The contribution of Gaussian amplitude
fluctuations can be obtained by fitting to the Aslamazov-Larkin DC
form.  Using the procedure of Gantmakher \cite{Gantmakher}, a
lower bound on $T_{c0}$ can be estimated as the lowest temperature
that does not cause an inflection point in the extracted effective
normal state resistance $R_N(T)$ as defined by the full expression
for the Aslamazov-Larkin fluctuation resistivity.  Within this analysis $2.28 K$ is the best lower bound on $T_{c0}$ and represents the temperature scale below which the
superconducting amplitude is relatively well defined. Lower values
of  $T_{c0}$ produce a kink in the extracted resistivity, where
for instance a $T_{c0}$ of $2.2K$ is clearly too low as seen in
Fig. \ref{AC1} (top).

In Fig. \ref{AC1} (bottom), we plot the superfluid stiffness vs. temperature as defined above.  We see a broad temperature region characterized by a gentle roll off of the
superfluid stiffness with increasing temperature.  Moreover, we observe a
distinct temperature where the superfluid stiffness acquires a
frequency dependence. Within the standard theory, such an
occurrence is indicative of the approach to a KTB transition. In
this model, the \textit{zero-frequency} superfluid stiffness is
renormalized discontinuously to zero at a temperature $T_{KTB}$
set by the superfluid stiffness itself at this temperature.  Above
$T_{KTB}$ the system still appears superconducting on short length
scales set by the separation between thermally generated free
vortices and therefore at finite frequencies we expect the
superfluid stiffness to approach zero \textit{continuously}.  As
$T_{KTB}$ is the temperature where vortices proliferate, in at
least moderate fugacity (a quantity related to the vortex core
energy) superconductors, the temperature
where the superfluid stiffness curves measured at different
frequencies deviate from each other can be identified as the approximate location of the $T_{KTB}$.  With our units of superfluid stiffness the predicted transition temperature is $
T_{KTB} = T_{\theta}/4 $, which is shown as a solid line in the
figure.  We point out here that although a frequency dependence as
such is seen in Fig. \ref{fig:dirtyBCS}, the stiffness where
$T_{\theta}$ acquires its frequency dependence is well above the
stiffness predicted to be critical value.  This point is expanded on in Ref. \cite{Crane06b}.

Note that any residual normal electron contribution should give an insignificant contribution to $\sigma_2$  \cite{Crane06,Crane06b}.   In principle there can also be a small
contribution to $\sigma_2$ from 'normal' electrons excited above the
gap.   However as these electrons are subject to essentially normal
state dissipative processes their contribution to $\sigma_2$ is very
small because the normal state scattering is so strong (the
material is amorphous with an extremely short mean free path).
These considerations should be particularly true at our relatively
low frequencies (low as compared to the normal state scattering
rate) in a highly disordered material with a very small normal
state $\sigma_1$ and large normal state scattering rate (1/$\tau$).  One can see this via the general theory of metallic conduction, within for instance the Drude model, where the magnitude of $\sigma_2$
becomes very small at measurement frequencies well below the
normal state scattering rate i.e. $\sigma_2(\omega)$ is generally very small
when $\omega \tau \ll 1$.  From an estimate of the mean free path in
this highly disordered system (6 nm) (from Steiner and Kapitulnik Ref. \cite{Steiner2} who used materials very similar to ours) and a reasonable order of magnitude guess of the Fermi velocity (1/200
the speed of light), one gets a very large normal state scattering
rate 250 THz.  At even our largest experimental
frequencies (106 GHz) $\omega \tau \ll 1$.  Using these numbers, we
can make a simple estimate of the order of magnitude of possible
'contamination' of our measured superfluid stiffness from above
gap electrons.  If ALL the charge carriers that participate in
pairing became subject to normal state dissipative processes they
would contribute a contamination signal T$_{cont}$ to the superfluid
stiffness, which can be estimated using the relation from the Drude model $T_{cont}  = T_{\theta}[T=0](\omega \tau)^2$.  This is a worst case scenario of the contribution to $\sigma_2$ from these
carriers and follows from the Drude formula and our definition of T$_\theta$. We find that even at our maximum frequency of 106 GHz, the largest possible contamination to our superfluid stiffness is
of the order of $2 \times 10^{-6}$ Kelvin which can be compared to our low
temperature value of 7.66 K. This is far below our sensitivity
level. It is a consequence of the large normal state scattering
and is one of the added benefits of using highly disordered
samples. In clean materials, in principle it is possible to have a
contribution from normal or above gap electrons. Note that even if
we have overestimated the scattering rate by a factor of a
hundred, this contamination signal becomes only of order  0.02K, which is
still well below our sensitivity.

\begin{figure}[htbp]
\begin{center}
\includegraphics[width=8cm,angle=0]{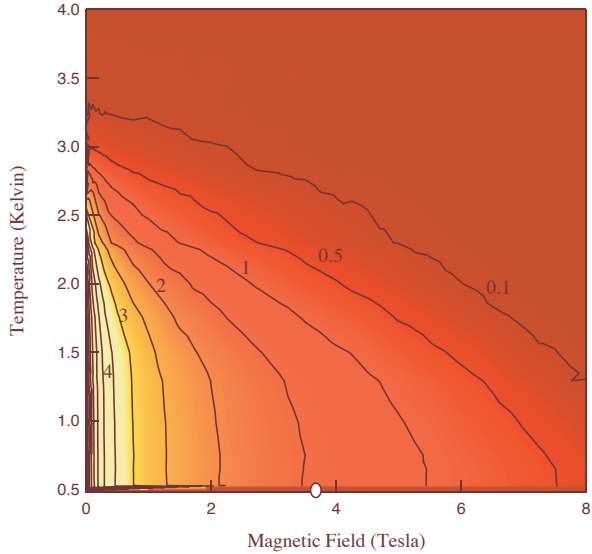}
\caption{ (left) Superfluid stiffness ($\propto \omega \sigma_2$) at 22 GHz given in units of degrees Kelvin.  The critical field, $H_{SIT}$, defined as the iso-resistance point in DC measurements, is shown as a white dot at $H=3.68 T$. Yellow indicates maximum.  Contours with markers are the detection limit for $T_{\theta}$ at specified frequencies. Black dots on the temperature axis denote $T_{amp}$ ($=T_{c0}$) and $T_{phase}$ ($=T_{KTB}$) which are temperatures signifying the freezing of amplitude and phase fluctuations, respectively.   $H_{SIT}$ appears as a black dot on the horizontal axis.  Open symbols represent data obtained from a small linear extrapolation beyond our maximum field of 8 Tesla. }
\label{AC2}
\end{center}
\end{figure}

Along the `quantum' axis, we can follow the superfluid stiffness signal across the critical point and into the insulating state.  In Fig. \ref{AC2} we display the field and temperature dependence of the generalized finite frequency superfluid stiffness measured at 22 GHz, $T_{\theta}$.   Similar plots can be made at our other measurement frequencies.    The finite-frequency superfluid stiffness falls quickly with increasing field, but remains finite above $H_{SIT}$ and well into the insulating regime to fields almost 3 times the critical field H$_{SIT} = 3.68 T$.  Here the critical field  has been defined by the iso-resistance point from DC measurements that were performed concurrently \cite{Crane06b}.  This is one of the first direct model-free measure of superconducting correlations on the insulating side of the 2D superconductor-insulator transition in an amorphous film.  It is the use of relatively high probing frequencies that allows us to resolve superconducting fluctuations into the insulating part of the phase diagram.

Our observation of a finite frequency superfluid stiffness at $H>H_{SIT}$ is not inconsistent with an insulating T=0 ground state. As alluded to above, our experiments are sensitive to superfluid \textit{fluctuations} because we probe the system on short time scales via an experimental frequency $\omega_{Exp}$ that is presumably high compared to an intrinsic order parameter fluctuation rate $\omega_{QC}$ close to the transition.  We note that at low temperatures and well into the insulating side of the phase diagram, the superfluid stiffness becomes temperature independent as $T \rightarrow 0$.  This shows that the observed effects are not thermally driven and is indicative of their intrinsic quantum mechanical nature.  Although we cannot rule out inhomogeneous superconducting patches \cite{Nandini}, we consider the large  ($10^6 \Omega/\Box$) and strongly diverging resistance of our samples at the highest fields and low temperatures to at least favor an interpretation of a material which is globally insulating.

These measurements give the first direct evidence for quantum superconducting fluctuations around an
insulating ground state and moreover, the unambiguous
evidence for a state of matter with localized Cooper pairs. The
finite frequency superfluid stiffness was an increasing function
of $\omega$ showing that superconductivity, although fluctuating
on longer length scales, was increasingly well defined on the
shorter length scales.   These measurements give further evidence for a novel insulating state which has collective properties in accord with the the DC measurements discussed in detail above.

\section{Summary}
\begin{figure}
\includegraphics[width=6.7cm,angle=0]{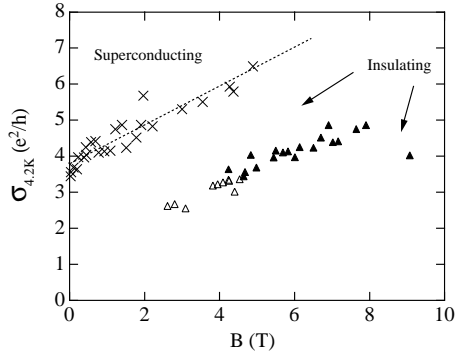}
\caption{Experimental phase diagram for a:InO superconducting thin-films. The crosses mark the $B_c$ of each sample at the SIT. The dashed line best fits the $B_c$ data. Empty and filled triangles mark the positions of the $B$-induced insulating peak for samples that, at $B$ = 0, are insulating and superconducting respectively.}
\label{Phase}
\end{figure}

To summarize, intriguing results from DC and AC transport measurements of thin films of a:InO that were driven through a SIT by the application of perpendicular $B$ are presented. The observations are discussed from a point of view that the results point to a possible relation between the conduction mechanisms in the superconducting and insulating phases in these disordered films. We summarize our findings in the form of an experimental phase diagram \cite{FisherPRL1}, shown in Fig. \ref{Phase}, plotted in the $B$-disorder plane. It is constructed based on our data from 11 different a:InO films at 33 different annealing stages. The normal-state conductivity of the films at $T$ = 4.2 K ($\sigma_{4.2K}$) is used as a measure of disorder. Conductivity at higher $T$'s could have equally been used with no essential change to the phase diagram. Two kinds of data points are plotted. The crosses mark the $B_c$ of each film and thus define the boundary between the superconducting and insulating phases. The dashed line is a best fit to the $B_c$ data. Extrapolation of this fit to $B$ = 0 gives a $\sigma_{4.2K}$ value of 3.8$\pm$0.1 $e^2/h$ ($e$ is the electronic charge and $h$ is Planck's constant). Triangles are used to identify the $B$-position of the insulating peak for samples that, at $B$ = 0, were insulating (empty triangles) or superconducting  (filled triangles). Our phase diagram is similar to the schematic phase diagram for a 2D superconductor suggested in \cite{Paalanen92}.

GS acknowledges support from NSF DMR-0847324 and University at Buffalo's IRDF grant. NPA is supported by NSF DMR-0847652.  The majority of the DC measurements were performed at the Weizmann Institute of Science under the guidance of Prof. Dan Shahar. Some measurements were performed at the National High Magnetic Field Laboratory which is supported by the NSF Cooperative Agreement No. DMR-0084173 and by the State of Florida.


\thebibliography{0}

\bibitem{Hebard90} A. F. Hebard and M. A. Paalanen, ``Magnetic-field-tuned superconductor-insulator transition in two-dimensional films", Phys. Rev. Lett. \textbf{65}, 927-930
(1990).

\bibitem{FisherPRL1}   M. P. A. Fisher, ``Quantum phase transitions
in disordered two-dimensional superconductors", Phys. Rev. Lett.
\textbf{65}, 923 (1990).

\bibitem{FisherPRL2} M. P. A. Fisher, G. Grinstein, and S. M.
Girvin, ``Presence of quantum diffusion in two dimensions:
Universal resistance at the superconductor-insulator transition",
Phys. Rev. Lett. \textbf{64}, 587 (1990).

\bibitem{Fink94} A.M. Finkel'stein, ``Suppression of superconductivity in homogenously disordered systems", Physica B \textbf{197}, 636 (1994).

\bibitem{Feigel1} M. V. Feigel'man, L. B. Ioffe, V. E. Kravtsov, and E. A. Yuzbashyan, ``Eigenfunction Fractality and Pseudogap State near the Superconductor-Insulator Transition", Phys. Rev. Lett. \textbf{98}, 027001 (2007).

\bibitem{Nandini} Amit Ghosal, Mohit Randeria, and Nandini Trivedi, Phys. Rev. Lett. \textbf{81}  3940 (1998).

\bibitem{Meir07} Yonathan Dubi, Yigal Meir and Yshai Avishai, ``Nature of the superconductor-insulator transition in disordered superconductors", Nature \textbf{449}, 876 (2007)

\bibitem{Markovic98} A. M. Goldman and N. Markovic, ``Superconductor-insulator transitions in the two-dimensional limit", Physics Today \textbf{51}, 39 (1998).

\bibitem{White86} A. E. White, R. C. Dynes, and J. P. Garno, ``Destruction of superconductivity in quench-condensed two-dimensional films", Phys. Rev. B \textbf{33}, 3549
(1986).

\bibitem{Yazdani95} Ali Yazdani and Aharon Kapitulnik, ``Superconducting-Insulating Transition in Two-Dimensional a-MoGe Thin Films", Phys. Rev. Lett. \textbf{74}, 3037
(1995).

\bibitem{Gant98} V.F. Gantmakher, M.V. Golubkov, V.T. Dolgopolov,
G.E. Tsydynzhapov, and A.A. Shashkin, ``Destruction of localized
electron pairs above the magnetic-field-driven
superconductor-insulator transition in amorphous InO films", JETP
Lett. \textbf{68}, 363 (1998).

\bibitem{Butko01} V. Y. Butko and P. W. Adams, ``Quantum metallicity in a two-dimensional insulator", Nature (London), \textbf{409}, 161 (2001).

\bibitem{Biel02} E. Bielejec and W. Wu, ``Field-Tuned Superconductor-Insulator Transition with and without Current Bias", Phys. Rev. Lett. \textbf{88}, 206802
(2002).

\bibitem{Murthy04} G. Sambandamurthy, L. W. Engel, A. Johansson,
and D. Shahar, ``Superconductivity-Related Insulating Behavior",
Phys. Rev. Lett. \textbf{92}, 107005 (2004).

\bibitem{Steiner2} Myles Steiner and Aharon Kapitulnik,
``Superconductivity in the insulating phase above the field-tuned
superconductor-insulator transition in disordered indium oxide
films", Physica C \textbf{422}, 16 (2005).

 \bibitem{Crane06} R. W. Crane, N. P. Armitage, A. Johansson, G. Sambandamurthy, D. Shahar, and G.
Gruner, ``Survival of superconducting correlations across the two-dimensional superconductor-insulator transition: A finite-frequency study ",
Phys. Rev. B \textbf{75}, 184530 (2007).

\bibitem{Valles07} M. D. Stewart, Jr., Aijun Yin, J. M. Xu, and James M. Valles, Jr.,  ``Superconducting pair correlations in an amorphous insulating nanohoneycomb film", Science \textbf{318}, 1273 (2007).

\bibitem{Baturina07} T.I. Baturina, A.Yu. Mironov, V.M. Vinokur, M.R. Baklanov, and C. Strunk, ``Localized Superconductivity in the Quantum-Critical Region of the Disorder-Driven Superconductor-Insulator Transition in TiN Thin Films", Phys. Rev. Lett. \textbf{99}, 257003 (2007).
 
 \bibitem{Sacepe08} B. Sacepe, C. Chapelier, T. I. Baturina, V. M. Vinokur, M. R. Baklanov, M. Sanquer, ``Disorder-induced inhomogeneities of the Superconducting State Close to the Superconductor-Insulator Transition", Phys. Rev. Lett. \textbf{101}, 157006 (2008).
 
\bibitem{Sondhi97} S. L. Sondhi, S. M. Girvin, J. P. Carini, and D. Shahar, ``Continuous quantum phase transitions", Rev. Mod. Phys. \textbf{69}, 315
(1997).

\bibitem{Sachdevbook} Subir Sachdev, ``Quantum Phase Transitions", Cambridge University
Press, (1999).

\bibitem{Chris02} C. Christiansen, L. M. Hernandez, and A. M.
Goldman, ``Evidence of Collective Charge Behavior in the Insulating
State of Ultrathin Films of Superconducting Metals", Phys. Rev.
Lett. \textbf{88}, 037004 (2002).

\bibitem{Valles92} J. M. Valles, Jr., R. C. Dynes, and J. P. Garno, ``Electron tunneling determination of the order-parameter amplitude at the superconductor-insulator transition in 2D", Phys. Rev. Lett. \textbf{69}, 3567 (1992).


\bibitem{Liu91} Y. Liu, K. A. McGreer, B. Nease, D. B. Haviland,
G. Martinez, J. W. Halley, and A. M. Goldman, ``Scaling of the
insulator-to-superconductor transition in ultrathin amorphous Bi
films", Phys. Rev. Lett. \textbf{67}, 2068 (1991).

\bibitem{Paalanen92} M. A. Paalanen, A. F. Hebard, and R. R. Ruel, ``Low-temperature insulating phases of uniformly disordered two-dimensional superconductors", Phys. Rev. Lett. \textbf{69}, 1604 (1992).

\bibitem{Kowal94} D. Kowal and Z. Ovadyahu, ``Disorder induced granularity in an amorphous superconductor", Solid State Commun. \textbf{90}, 783
(1994).

\bibitem{Meyer01} J. S. Meyer and B. D. Simons, ``Gap fluctuations in Inhomogeneous superconductors", Phys. Rev. B \textbf{64}, 134516 (2001).
  
 \bibitem{Gant00} V. F. Gantmakher, M. V. Golubkov, V. T. Dolgopolov, G. E. Tsydynzhapov and A. A. Shashkin,``Scaling Analysis of the Magnetic FieldÐTuned Quantum
Transition in Superconducting Amorphous InÐO Films", JETP Letters, \textbf{71}, 160 (2000).
  
\bibitem{Gant95} V. F. Gantmakher and M. V. Golubkov , ``Superconductivity and negative magnetoresistance in amorphous InO films", JETP Letters, \textbf{61}, 606 (1995).
  
 \bibitem{Phillips03} P.Phillips and D. Dalidovich, ``The Elusive Bose Metal", Science, \textbf{302}, 243 (2003).
  
 \bibitem{Ephron96} D. Ephron, A. Yazdani, A. Kapitulnik and M. R. Beaseley, ``Observation of Quantum Dissipation in the Vortex State of a Highly Disordered Superconducting Thin Film", Phys. Rev. Lett. \textbf{76}, 1529 (1996).
  
 \bibitem{Chervenak00} J. A. Chervenak and J. M. Valles Jr., ``Absence of a zero-temperature vortex solid phase in strongly disordered superconducting Bi films", Phys. Rev. B \textbf{61}, R9245 (2000).
 
\bibitem{Gorkov89} L. P. Gorkov and G. Gruner (Eds),  \textit{Charge density waves in Solids}, North-Holland, Amsterdam (1989). 

\bibitem{Goldman90} V. J. Goldman, M. Santos, M. Shayegan and J. E. Cunningham, ``Evidence for two-dimentional quantum Wigner crystal", Phys. Rev. Lett. \textbf{65}, 2189,1990.

\bibitem{Jiang91} H. W. Jiang, H. L. Stormer, D. C. Tsui, L. N. Pfeiffer and K. W. West, ``Magnetotransport studies of the insulating phase around $\nu$=1/5 Landau-level filling", Phys. Rev. B \textbf{44}, 8107 (1991).
 
\bibitem{Williams91} F. I. B. Williams, P. A. Wright, R. G. Clark, E. Y. Andrei, G. Deville, D. C. Glattli, O. Probst, B. Etienne, C. Dorin, C. T. Foxon, and J. J. Harris,   ``Conduction threshold and pinning frequency of magnetically induced Wigner solid", Phys. Rev. Lett. \textbf{66}, 3285 (1991).

\bibitem{Ovadia09} M. Ovadia, B. SacŽpŽ, and D. Shahar, ``Electron-Phonon Decoupling in Disordered Insulators",  Phys. Rev. Lett. \textbf{102}, 176802 (2009).

\bibitem{Altshuler09} Boris L. Altshuler, Vladimir E. Kravtsov, Igor V. Lerner, and Igor L. Aleiner, ``Jumps in Current-Voltage Characteristics in Disordered Films", Phys. Rev. Lett.  \textbf{102}, 176803 (2009).

\bibitem{Vinokur10} A. Petkovic, N. M. Chtchelkatchev, T. I. Baturina, and V. M. Vinokur, ``Out-of-Equilibrium Heating of Electron Liquid: Fermionic and Bosonic Temperatures", Phys. Rev. Lett. \textbf{105}, 187003 (2010).

\bibitem{Blatter94a} G. Blatter, M. V. Feigel'man, V. B. Geshkenbein, A. I. Larkin and V. M. Vinokur, ``Vortices in high-temperature superconductors", Rev. Mod. Phys. \textbf{66}, 1125 (1994).

\bibitem{Palstra88} T. T. M. Palstra, B. Batlogg, L. F. Schneemeyer, and J. V. Waszczak, ``Thermally Activated Dissipation in Bi$_{2.2}$Sr$_2$Ca$_{0.8}$Cu$_2$O$_{8+\delta}$", Phys. Rev. Lett. \textbf{61}, 1662 (1988)

\bibitem{Inui89} M. Inui, P. B. Littlewood, and S. N. Coppersmith, ``Pinning and thermal fluctuations of a flux line in high-temperature superconductors", Phys. Rev. Lett. \textbf{63}, 2421 (1989).

\bibitem{White93} W. R. White, A. Kapitulnik, and M. R. Beasley, ``Collective vortex motion in a-MoGe superconducting thin films", Phys. Rev. Lett. \textbf{70}, 670 (1993). 

\bibitem{Chervenak96} J. A. Chervenak and J. M. Valles, Jr., ``Evidence for a quantum-vortex-liquid regime in ultrathin superconducting films", Phys. Rev. B \textbf{54}, R15649 (1996). 

\bibitem{Shahar98} D.Shahar, M.Hilke, C.C. Li, D.C. Tsui, S.L. Sondhi, J.E. Cunningham, M. Razeghi, ``A New Transport Regime in the Quantum Hall Effect", Solid State Comm. \textbf{107}, 19 (1998).
  
\bibitem{Crane06b} R. W. Crane, N. P. Armitage, A. Johansson, G. Sambandamurthy, D. Shahar, and G.
Gruner, ``Fluctuations, dissipation, and nonuniversal superfluid jumps in two-dimensional superconductors", Phys. Rev. B \textbf{75}, 094506 (2007).

\bibitem{Berez} V.L. Berezinskii, ``Destruction of long-range order in one-dimensional and 2-dimensional systems processing a continuous symmetry group. 1. Classical systems", Sov. Phys. JETP \textbf{32}, 493 (1971);
V.L. Berezinskii, ``Destruction of long-range order in one-dimensional and 2-dimensional systems processing a continuous symmetry group. 2. Quantum systems" Sov. Phys. JETP \textbf{34},
610 (1972)

\bibitem{KToriginal} M. Kosterlitz and D. Thouless, ``Ordering, metastability and phase transitions in two-dimensional systems", J. Phys. C \textbf{6}, 1181 (1973).

\bibitem{KTBhelium}    D. J. Bishop and J. D. Reppy, ``Study of the Superfluid Transition in Two-Dimensional 4He Films", Phys. Rev. Lett. \textbf{40}, 1727–1730
(1978).

\bibitem{Hebard1985} A. F. Hebard and M. A. Paalanen, ``Diverging Characteristic Lengths at Critical Disorder in Thin-Film Superconductors", Phys. Rev. Lett. \textbf{54},
2155 (1985).

\bibitem{Halperin} B.I. Halperin and D.R. Nelson, ``Resistive transition in superconducting films", J. Low Temp. Phys. \textbf{36}, 599
(1979).

\bibitem{KTdynamics} V. Ambegaokar, B. I. Halperin, David R. Nelson, and Eric D.
Siggia, ``Dynamics of superfluid films", Phys. Rev. B \textbf{21},
1806–1826 (1980).


\bibitem{Hebard80a}  A. F. Hebard and A. T. Fiory, Phys. Rev. Lett. \textbf{44}, 291 - 294 (1980).

\bibitem{Fiory83a} A. T. Fiory, A. F. Hebard, and W. I. Glaberson, Phys. Rev. B \textbf{28}, 5075 - 5087 (1983).

\bibitem{Adams87a} P. W. Adams and W. I. Glaberson,  Phys. Rev. B \textbf{35}, 4633 - 4652 (1987)

\bibitem{Corson} J. Corson, R. Mallozzi, J. Orenstein, J. N.
Eckstein  and I. Bozovic, ``Vanishing of phase coherence in
underdoped Bi$_2$Sr$_2$CaCu$_2$O$_{8+\delta}$", Nature
\textbf{398}, 221 (1999).

\bibitem{Gantmakher} V. F. Gantmakher and M.V. Golubkov, JETP Letters
\textbf{73}, 131 (2001).


\bibitem{KotzlerBrandt1} E. H. Brandt, Phys. Rev. Lett. \textbf{71}, 2821
(1993).

\bibitem{KotzlerBrandt3} J. K\"{o}tzler, G. Nakielski, M. Baumann, R. Behr, and F. Goerke, E. H. Brandt,Phys. Rev. B \textbf{50}, 3384 (1994).

\bibitem{Redbook} O. Klein, S. Donovan, M. Dressel, G.
Gr\"{u}ner, \textit{Int. J. Infrared Millim. Waves} \textbf{14},
2423 (1993).

\bibitem{Peligrad} D.-N. Peligrad, B. Nebendahl, M. Mehring, A. Dulcic, M. Pozek, and D. Paar, Phys. Rev. B \textbf{64}, 224504
(2001).

\bibitem{Waldron} R. A. Waldron, \textit{The Theory of Waveguides and Cavities}, (Maclaren \& Sons Ltd., London, 1967).

\bibitem{Peligrad2} D.-N. Peligrad, B. Nebendahl, C. Kessler, M. Mehring, A. Dulcic, M. Pozek,  D. Paar, Phys. Rev. B \textbf{58}, 11652 (1998).

\endthebibliography
\end{document}